\documentclass[conference]{IEEEtran}
\IEEEoverridecommandlockouts
\usepackage{ifpdf}
\usepackage{cite}
\usepackage[pdftex]{graphicx}
\usepackage{amsmath}
\usepackage{algorithmic}
\usepackage{array}
\usepackage{caption}
\usepackage[caption=false,font=footnotesize]{subfig}

\usepackage{fixltx2e}
\usepackage{stfloats}
\usepackage{url}
\usepackage{amsthm}
\usepackage{amsmath}
\usepackage{amsfonts}
\usepackage{amssymb}
\usepackage[T1]{fontenc} 
\usepackage{amsmath}
\usepackage[cmintegrals]{newtxmath}
\usepackage{bm} 
\usepackage[all]{xy}
\usepackage{enumitem}
\usepackage{float}
\usepackage{amsmath}
\usepackage[usenames, dvipsnames]{color}
\usepackage{ifpdf}
\usepackage{cite}

\usepackage{amsmath}
\usepackage{algorithmic}
\usepackage{array}
\usepackage{mathtools}

\makeatletter
\def\widebreve{\mathpalette\wide@breve}
\def\wide@breve#1#2{\sbox\z@{$#1#2$}%
\mathop{\vbox{\m@th\ialign{##\crcr
\kern0.08em\brevefill#1{0.8\wd\z@}\crcr\noalign{\nointerlineskip}%
$\hss#1#2\hss$\crcr}}}\limits}
\def\brevefill#1#2{$\m@th\sbox\tw@{$#1($}%
 \hss\resizebox{#2}{\wd\tw@}{\rotatebox[origin=c]{90}{\upshape(}}\hss$}
\makeatletter

\usepackage{url}
\usepackage{mathtools}
\usepackage{amsthm}
\usepackage{amsmath}
\usepackage{amsfonts}
\usepackage{amssymb}
\usepackage[T1]{fontenc} 
\usepackage{amsmath}
\usepackage[cmintegrals]{newtxmath}
\usepackage{bm} 
\usepackage[all]{xy}
\usepackage{enumitem}
\usepackage{float}
\usepackage{amsmath}
\usepackage[dvipsnames]{xcolor}
\usepackage{multirow}
\usepackage[T1]{fontenc}
\usepackage{mwe} 
\usepackage{times}
\usepackage{wasysym}
\usepackage{epsfig}
\usepackage{multirow}

\usepackage{tabularx}
\usepackage{graphicx}
\usepackage{color}
\usepackage{xspace}
\usepackage{thumbpdf}
\usepackage{listings}
\usepackage{verbatim}
\usepackage{outlines}
\usepackage{textgreek}
\usepackage{booktabs}
\usepackage{amsfonts}
\usepackage{colortbl}
\usepackage{multicol}
\usepackage{multirow}
\usepackage{balance}
\usepackage{makecell}
\usepackage{array}
\usepackage[normalem]{ulem}
\usepackage{mathtools,amssymb}
\usepackage{float} 
\usepackage{booktabs} 
\usepackage{wrapfig}
\usepackage{subfloat}

\usepackage{diagbox}

\theoremstyle{definition}

\newcommand{\no}{\nonumber}

\begin{document}
	\title{Leveraging Prior Knowledge Asymmetries in the Design of Location Privacy-Preserving Mechanisms}
 	\author{\IEEEauthorblockN{Nazanin Takbiri}
 \IEEEauthorblockA{Electrical and\\Computer Engineering\\
 	UMass-Amherst\\
 	ntakbiri@umass.edu}
 \and
 \IEEEauthorblockN{Virat Shejwalkar}
 \IEEEauthorblockA{Information and \\Computer Sciences\\
 		UMass-Amherst \\
 	vshejwalkar@cs.umass.edu }
 \and
 \IEEEauthorblockN{Amir Houmansadr}
 \IEEEauthorblockA{Information and \\Computer Sciences\\
 		UMass-Amherst \\
 	amir@cs.umass.edu}
 \and
 \IEEEauthorblockN{Dennis L. Goeckel}
 \IEEEauthorblockA{Electrical and\\Computer Engineering\\
 	UMass-Amherst\\
 	goeckel@ecs.umass.edu}
 \and
 \IEEEauthorblockN{Hossein Pishro-Nik}
 \IEEEauthorblockA{Electrical and\\Computer Engineering\\
 	UMass-Amherst\\
 	pishro@ecs.umass.edu\thanks{This work was supported by National Science Foundation under grants CCF--1421957 and CNS--1739462.}
}
}

	\maketitle

\begin{abstract}
	
The prevalence of mobile devices and Location-Based Services (LBS) necessitate the study of Location Privacy-Preserving Mechanisms (LPPM). However, LPPMs reduce the utility of LBS due to the noise they add to users' locations. Here, we consider the remapping technique, which presumes the adversary has a perfect statistical model for the user location. We consider this assumption and show that under practical assumptions on the adversary's knowledge, the remapping technique leaks privacy not only about the true location data, but also about the statistical model. Finally, we introduce a novel solution called  ``Randomized Remapping'' as a countermeasure.

\end{abstract}



\section{Introduction}
\label{intro}

Mobile devices, ranging from smart phones to connected automobiles, provide ubiquitous communication and offer a wide spectrum of location-based services (LBS), such as ride sharing, navigation, dining recommendations,
and accident warnings. LBS collect large amounts of users' location data to tailor the service provided to each user's specific needs.
To address the significant threat to user privacy due to location data sharing~\cite{FTC2015}, multiple location privacy-preserving mechanisms (LPPMs) are proposed in the literature.
Location data obfuscation is the main theme of all LPPMs, which enhances privacy by using misleading, false, or ambiguous information~\cite{gruteser2003anonymous, bordenabe2014optimal, obf_1, obf_3, shokri2012protecting,Nazanin_IT,ISIT18-longversion}. 
However, obfuscation degrades system utility to enhance privacy~\cite{shokri2012protecting,uti_1}.

Recently, a remapping technique was proposed to improve the utility of LPPMs without compromising privacy~\cite{remap1}.
Remapping addresses privacy against an adversary with perfect knowledge of the prior distribution of a user's location data.
Instead of releasing the obfuscated location, remapping releases an estimated location that the adversary would infer anyway.
Therefore, remapping improves utility as the estimated location is closer to the true location than the obfuscated location.

We model remapping as a general utility improvement technique for releasing not just location data but any type of data, e.g., IoT application data.
Therefore, we consider Gaussian distributed private data whose privacy is protected by adding carefully calibrated Gaussian noise to it.
Gaussian distributed data has been long considered in various domains, e.g., sensor networks~\cite{xiao2005scheme,vempaty2012localization,ribeiro2006bandwidth} and distributed consensus~\cite{wagner2004resilient}, as a promising substitute to the real data. 
Therefore, our analysis of the remapping technique for Gaussian distributed data generalizes to all such domains. In addition, modeling the probability of a user’s check-in at a location based on a Multi-center Gaussian Model is being employed widely~\cite{gau1,gau2,gau3,gau4,gau5}. A significant characteristic of check-in locations is that they are usually located around several centers, and the probability of a user visiting a location is inversely proportional to the distance from its nearest center; thus, users' check-in behavior can be modeled with a Gaussian distribution. Our Gaussian model could thus be used within a given region of the multi-center Gaussian model.

Here, we consider a friend (e.g., an IoT application on a smart phone) without any prior statistical information about user behavior and an adversary with statistical information about the user's behavior. This may occur, for example, when each intended recipient is either naive or only looking at a single datum or a small set of data from the user, whereas the adversary is sophisticated and has access to the user's data across a large time period.

In such a case, the adversary can use their statistical advantage to obtain a better estimate of the user's data than the friend. Remapping recognizes this fact and reveals a more accurate version of the data that the adversary would have been able to obtain anyway using her statistical advantage. Thus, remapping technique does not incur privacy loss, but improves accuracy for the user. Hence, by recognizing this asymmetry in prior knowledge, utility has been improved without privacy loss versus standard obfuscation; a simple example of the mechanism is demonstrated in Section \ref{framework}. Not surprisingly, this approach has garnered a growing amount of interest in the privacy community~\cite{Palia2017OptimizingNL,rem2,rem3,rem4,rem5}, hence motivating a more fundamental analysis.

Note that the classical remapping technique implicitly assumes that the sophisticated adversary has knowledge of the exact statistical model of the user data.
However, in practice this is not the case, as the adversary's knowledge is a noisy version of the exact statistical model of the user data due to multiple reasons, e.g., the user not reporting some of her data, a limited history, or the reported data being noisy due to a glitch or noise in its transmission.
We explore the remapping technique from the lens of this practical setting where the adversary has an \emph{imperfect knowledge} of the statistical model of the user's data.
We detail our contributions next.

\hspace{-0.14 in}{\textbf{Contributions:}}
In this paper, we take the first information-theoretic look at this remapping technique. We introduce a simple information-theoretic model to explain how the utility can be improved without a loss in privacy. Next, we employ our model to explore important aspects of remapping that have not been considered before. 
As acknowledged briefly in~\cite{remap1}, a risk of remapping is that it relies critically on accurate knowledge of the adversary's statistical model. 
In particular, if the adversary does not have accurate statistical information and the user employs remapping, we show that privacy leaks in two separate ways: (i) the adversary obtains a more accurate version of the data than they would have had without remapping, and (ii) the adversary is able to improve their statistical knowledge of the users' data beyond what they would have been able to do without remapping. Interestingly, we will see that the second type of leakage is increased if the obfuscation noise is increased. We provide the first analysis of the loss of privacy due to each of these factors. After analyzing the loss in privacy under standard remapping~\cite{remap1}, we next turn to countermeasures. We introduce a random remapping algorithm, where data points are independently remapped with some probability. For a given utility for the intended recipient, this approach greatly complicates model improvement at the adversary versus deterministic remapping approaches, thus improving the privacy-utility trade-off.

The rest of the paper is organized as follows. In Section \ref{fram}, we present the framework: system model, metrics, and definitions. We next demonstrate in Section \ref{case1} how the current remapping technique proposed by~\cite{remap1} increases utility while satisfying the same level of privacy if the adversary has the \textit{perfect prior} (i.e. a perfect statistical model of the user's data). In Section \ref{case2}, we quantify information leakage caused by the mentioned remapping technique~\cite{remap1} if the adversary does not have a perfect prior. In particular, we show that leakage about the distribution of the true data is a serious issue. Motivated by this, in Section \ref{rand} we propose a new method called "\textit{Randomized Remapping}" to improve the privacy for a given utility in this situation. This method provides a trade-off between leakage of the distribution of the true data and the utility. In Section \ref{con}, we draw conclusions from the work and present ideas for the future.

Due to space limitations, detailed derivations, additional discussion, and more references are provided in the long version of the paper~\cite{ISIT2019_longversion}.

\section{System Model and Metrics}\label{framework}
\label{fram}
\begin{figure}[t]
	\centering
	\includegraphics[width = 1\linewidth]{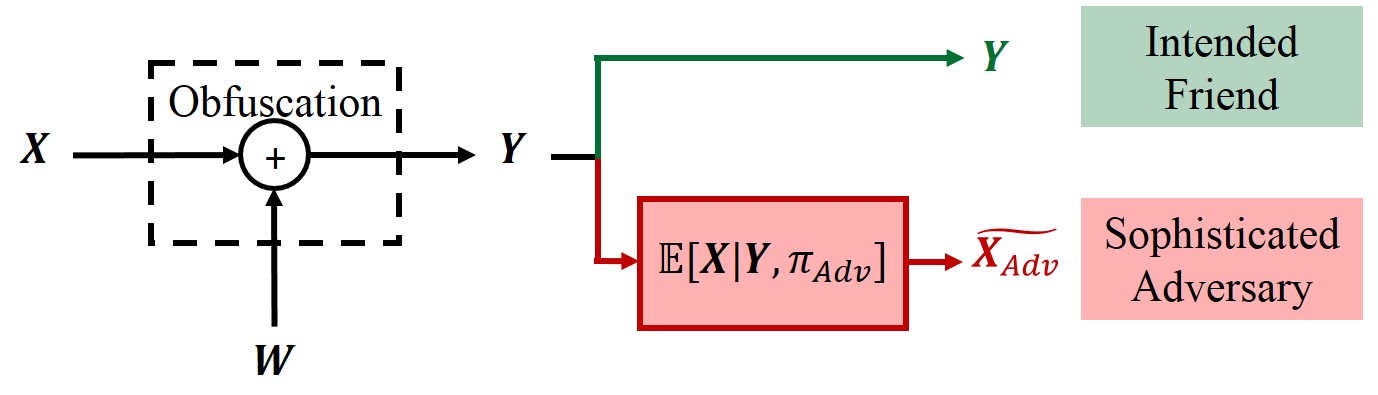}
	\caption{System Model: Case where additive obfuscation (without remapping) is applied to the user's location. The (naive) intended friend does not have a prior distribution for ${X}$ and hence employs ${Y}$ for the user's locations. A sophisticated adversary, who possesses a prior distribution for ${X}$, can use this prior to obtain a better estimate of the user's location.}
	\label{fig:model}
\end{figure}

\begin{figure}[t]
	\centering
	\includegraphics[width = 1\linewidth]{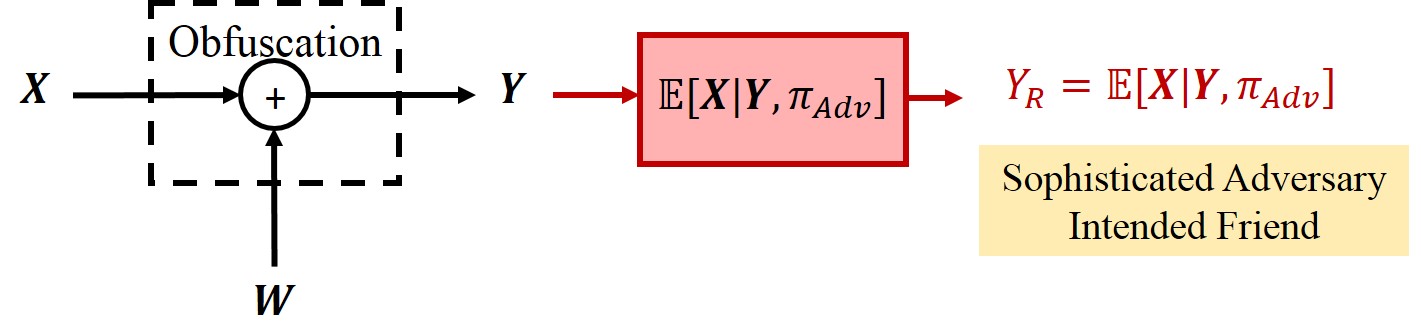}
	\caption{
	Remapping: ${X}$ is the user's true location, ${W}$ is the amount of noise added through the obfuscation process, ${Y}$ is the noisy reported location after applying obfuscation, and ${Y}_R$ is the remapped location which is the best possible estimate of the adversary according to perfect prior knowledge about the user.
	}
	\label{fig:remap2}
\end{figure}

Consider a system where a user generates a location $X$ which should be protected from a potential adversary. 
To preserve the privacy of the user's true location, the obfuscated location is obtained by adding noise $W$ to $X$.
In other words, the reported noisy version of location $({Y})$ is obtained as ${Y}={X}+{W}$.
As shown in Figure~\ref{fig:model}, there exists an ``intended'' friend (e.g., an LBS) who does not have prior statistical knowledge about the user behavior, and a ``sophisticated'' adversary who has knowledge about the prior behavior of the user $(\pi_{Adv})$. The adversary observes the noisy reported location ${Y}$ and uses it to find the estimate $\widetilde{{X}}_{Adv}$, which denotes the estimate of the adversary given their observed location $({Y})$ and their knowledge of the prior about the user $(\pi_{Adv})$ as
 $\widetilde{{X}}_{Adv}=\mathbb{E}\left[{X}|{Y}, \pi_{Adv}\right]$.
 As a result, there exist asymmetries in knowledge and/or sophistication between the intended friend and the adversary. The remapping technique, which is introduced by Chatzikokolakis et al.~\cite{remap1}, exploits these asymmetries to publish a more accurate version of the location that the sophisticated adversary would have been able to obtain anyway. 
As shown in Figure~\ref{fig:remap2}, each reported location is remapped into the best possible location according to the perfect prior information of the adversary.

\hspace{-0.14 in}\textbf{{Location Data Model:}}
We adopt a Gaussian model. User traces are assumed to be independent and identically distributed (i.i.d.) Gaussian series, and each data location is drawn from a normal distribution with mean $\mu$ and variance $\sigma^2_s$, $X(k)\sim \mathcal{N}\left(\mu, \sigma^2_s\right)$. We also assume there exists some underlying prior for the distribution of the mean ($\mu$); we also take this to be Gaussian, and hence assume $\mu \sim \mathcal{N}\left(0, \sigma^2_\mu\right)$.

\hspace{-0.14 in}\textbf{{Obfuscation Mechanism:}}
The obfuscated location is obtained by passing the data location through an additive white Gaussian noise (AWGN) channel~\cite{noise1,noise2,noise3}. Hence, ${Y}$, the reported location of the user, is the sum of the true location, ${X}$, and the noise, ${W}$, where $W$ is drawn from a zero-mean normal distribution with variance equal to $\sigma^2_w$. Thus, we have
\begin{align}
\no Y=X+W \sim \mathcal{N}\left(\mu, \sigma^2_s+\sigma^2_w\right).
\end{align}

\hspace{-0.14 in}\textbf{{Sophisticated Adversary Model:}}
The adversary logs the user's locations over time to generate a prior about the behavior of the user and performs an inference attack to estimate the best possible location given this generated prior.
Note that the remapping literature~\cite{remap1} has considered a \textit{perfect prior} for the adversary. In reality, the adversary, however strong, does not have an infinite time history of user's data or  have exact knowledge of the user's whereabouts, so she cannot build the perfect prior. In this paper, different adversarial settings have been considered: in Section~\ref{case1}, we assume an adversary with a perfect prior, and in Section~\ref{case2}, we assume an adversary with an imperfect prior. It is critical to note that the adversary knows the mechanism of the obfuscation, but she does not know the exact value of the noise which will be added during obfuscation and does not have any auxiliary information or side information about the user's location.

\hspace{-0.14 in}\textbf{{Remapping Mechanism:}}
In the absence of remapping, and given the \textit{perfect prior} for the adversary, the adversary can estimate ${X}$ using the reported noisy version of the location $({Y})$ as:
\begin{align}
{{Y}_{R}}=\mathbb{E}\left[{X}|{Y}, \mu\right]& = \frac{\sigma^2_w}{\sigma^2_s+\sigma^2_w}\mu+\frac{\sigma^2_s}{\sigma^2_s+\sigma^2_w}{Y},\ \
\label{X_R}
\end{align}
where ${{Y}_R}$ is the estimate of the adversary given the observed location $({Y})$ and perfect knowledge of the prior $(\pi_{Adv})$.
Remapping simply notes that, since the adversary obtains ${Y}_{R}$ anyway (as shown in Figure~\ref{fig:remap2}), we might as well provide it to the applications to improve the utility~\cite{remap1}.

\hspace{-0.14 in}\textbf{{Metrics:}}
In this paper, the mean squared error (MSE) is employed as a metric to quantify both utility degradation and privacy. In this paper, ``$\mathcal{U}$'' denotes the MSE of the intended application/friend which quantifies utility degradation. In addition, ``$\mathcal{P}$'' denotes the MSE of the adversary about the true location, and ``$\grave{\mathcal{P}}$'' denotes the MSE of the adversary about the statistical model. Note that both ``$\mathcal{P}$'' and ``$\grave{\mathcal{P}}$'' quantify the level of privacy.

\section{Case $1$: Adversary with Perfect Knowledge}
\label{case1}
In this section, we assume the adversary knows the exact statistical distribution of the user data, which for our model means the exact value of the mean $(\mu)$.

\subsection{Without Remapping}
\label{case1_NR}
Without remapping, the user's intended friend, who does not have any knowledge of the statistical model for the user's data, observes only the noisy location $(Y)$. The utility degradation is:
\begin{align} \mathcal{U}_{NR}^{(I)}&=\mathbb{E}\left[\left({\widetilde{X}_{App}}-X\right)^2\right]=\mathbb{E}\left[\left(Y-X\right)^2\right]=\sigma^2_w.\ \
\label{case1_NR_U}
\end{align}
In comparison to the user's friend, the sophisticated adversary obtains
 $\widetilde{X}_{Adv}=\mathbb{E}\left[X|Y,\mu\right]$. 
Thus, $\mathcal{P}_{NR}^{(I)}$ which quantifies the level of privacy is calculated as:
\begin{align}
\mathcal{P}_{NR}^{(I)}&=\mathbb{E}\left[\left({\widetilde{X}_{Adv}}-X\right)^2\right]=\mathbb{E}\left[\left({Y_R}-X\right)^2\right]=\frac{\sigma^2_w \sigma^2_s}{\sigma^2_s+\sigma^2_w}.\ \
\label{case1_NR_L}
\end{align}
\subsection{With Remapping}
\label{case1_R}
In this case, both the adversary and the user's friend observe the same reported location, $\widetilde{X}_{Adv}=\widetilde{X}_{App}=Y_R=\mathbb{E}\left[X|Y,\mu\right]$. Now, the MSE of the adversary and the MSE of the application are equal:
\begin{align}
\mathcal{U}_{R}^{(I)}=\mathcal{P}_{R}^{(I)}=\mathbb{E}\left[\left({Y_R}-X\right)^2\right]=\frac{\sigma^2_w \sigma^2_s}{\sigma^2_s+\sigma^2_w}.\ \
\label{case1_R_U}
\end{align}

Since the intended friend/application is oblivious to the prior statistical knowledge about the user behavior, the MSE of the adversary is always smaller than or equal to the MSE of the application $(\mathcal{P} \leq \mathcal{U})$. Thus, we can conclude that the remapping technique provides the best utility among techniques satisfying the same level of privacy under the assumption that the adversary has perfect knowledge of the statistical model for the user data~\cite{randomizedresponse}.

\section{Case $2$: Adversary with Imperfect Knowledge}
\label{case2}
Here, we assume the adversary has a noisy version of the prior information, as might be obtained from a learning set of limited length. Specifically, the adversary has $\check{\mu} =\mu +E$, where $E$ has a zero-mean normal distribution with variance equal to $\sigma^2_e$, as would be the case if $\check{\mu}$ were the minimum mean square estimate (MMSE) based on prior observations with additive Gaussian obfuscation. We consider not only the leakage of the true location $({X})$ but also the leakage of the distribution of the true location $(\mu)$, which is a serious issue, as such leakage would improve future estimates of the adversary.
 
\subsection{Without Remapping}
If remapping is not employed, the user's intended friend observes the reported location $(Y)$.
Thus, the utility is quantified as:
\begin{align}
 \mathcal{U}^{(II)}_{NR}&=\mathbb{E}\left[\left(\widetilde{X}_{App}-X\right)^2\right]=\mathbb{E}\left[\left(Y-X\right)^2\right]=\sigma^2_w.\ \
\label{NR_true_U_m}
\end{align}

In contrast, the sophisticated adversary uses both $Y=\mu+S +W$ and $\check{\mu} =\mu +E$ to improve knowledge not only about the true location $(X)$ but also about the distribution of the true location $(\mu)$. Now, $\grave{\mathcal{P}}^{(II)}_{NR}$ which quantifies the MSE of the adversary about the distribution of the true location $(\mu)$ is:

\begin{align}
\grave{\mathcal{P}}^{(II)}_{NR}=\mathbb{E}\left[\left(\widetilde{\mu}_{Adv}-\mu\right)^2\right]=\frac{\sigma_e^2\sigma_\mu^2\left(\sigma^2_s+\sigma^2_w\right)}{\left(\sigma^2_\mu+\sigma^2_e\right)\left(\sigma^2_s+\sigma^2_w\right)+\sigma^2_e\sigma^2_\mu}.\ \
\label{NR_distribution_L_two}
\end{align}
and $\mathcal{P}^{(II)}_{NR}$ which quantifies the MSE of the adversary about the true location $(X)$ is:
\begin{align}
\no &\mathcal{P}^{(II)}_{NR}=\mathbb{E}\left[\left(\widetilde{X}_{Adv}-X\right)^2\right]\\
 &=\frac{\sigma^2_s\sigma^2_w}{\sigma^2_s+\sigma^2_w}+\frac{\sigma^4_w\sigma^2_e\sigma^2_\mu}{\left(\sigma^2_s+\sigma^2_w\right)\left(\left(\sigma^2_\mu+\sigma^2_e\right)\left(\sigma^2_s+\sigma^2_w\right)+\sigma^2_e\sigma^2_\mu\right)}.\ \
\label{NR_true_L_two}
\end{align}


\subsection{With Remapping}
The user's friend observes the remapped location, so the utility of the system is:
\begin{align}
\mathcal{U}^{(II)}_{R}&=\mathbb{E}\left[\left({\widetilde{X}_{App}}-X\right)^2\right]=\mathbb{E}\left[\left({Y_R}-X\right)^2\right]=\frac{\sigma^2_s \sigma^2_w}{\sigma^2_s+\sigma^2_w}.\ \
\label{R_true_U_m}
\end{align}
However, the adversary observes not only $Y_R$, but also $\check{\mu}$, and uses both of these observations to estimate $\widetilde{\mu}_{Adv}$ and $\widetilde{X}_{Adv}$. 
Now, the MSE of the adversary about the distribution of the true location $(\mu)$ is:
\begin{align}
\grave{\mathcal{P}}^{(II)}_{R}&=\mathbb{E}\left[\left(\widetilde{\mu}_{Adv}-\mu\right)^2\right]=\frac{\sigma^4_s\sigma_e^2\sigma_\mu^2}{\sigma^4_s\left(\sigma^2_\mu+\sigma^2_e\right)+\sigma^2_e\sigma^2_\mu\left(\sigma^2_s+\sigma^2_w\right)},\ \
\label{R_distribution_L_two}
\end{align}
and the MSE of the adversary about the true location $(X)$ is:
\begin{align}
\no \mathcal{P}^{(II)}_{R}=&\mathbb{E}\left[\left(\widetilde{X}_{Adv}-X\right)^2\right]=\frac{\sigma^2_s\sigma^2_w}{\sigma^2_s+\sigma^2_w}.
\label{R_true_L_two}
\end{align}

Numerical results demonstrating what can be learned from these expression will be presented in Section~\ref{rand}.

\subsection{Discussion: Leakage of the Statistical Model}
From (\ref{R_distribution_L_two}), we can conclude that increasing the obfuscation noise, somewhat surprisingly, increases the leakage about the distribution of the true location $(\mu)$ when remapping is employed. Note that ${Y}_R=\mathbb{E}\left[{X}|{Y}, \check {\mu}\right]$ depends on two parameters: 1) $\check{\mu}=\mu+E$ and 2) ${Y}={X}+{W}$; thus, if we increase the obfuscation noise by increasing $\sigma^2_w$, ${Y}_R$ relies less on ${Y}$ and more on $\check{\mu}$. Now in the extreme case, where $\sigma^2_w$ goes to infinity, the observed location $({Y})$ is useless and, as a result, ${Y}_R=\mathbb{E}\left[{X}|{Y}, \check{\mu}\right]=\mu$. Hence, remapping technique leaks complete information about the statistical model $(\mu)$ as $\sigma^2_w$ goes to infinity.

\section{Randomized Remapping and Numerical Results}
\label{rand}
\begin{figure*}[t]
	\centering
	\subfloat[The MSE of the adversary about the statistical model $(\mu)$]{

		\includegraphics[width=0.9\columnwidth]{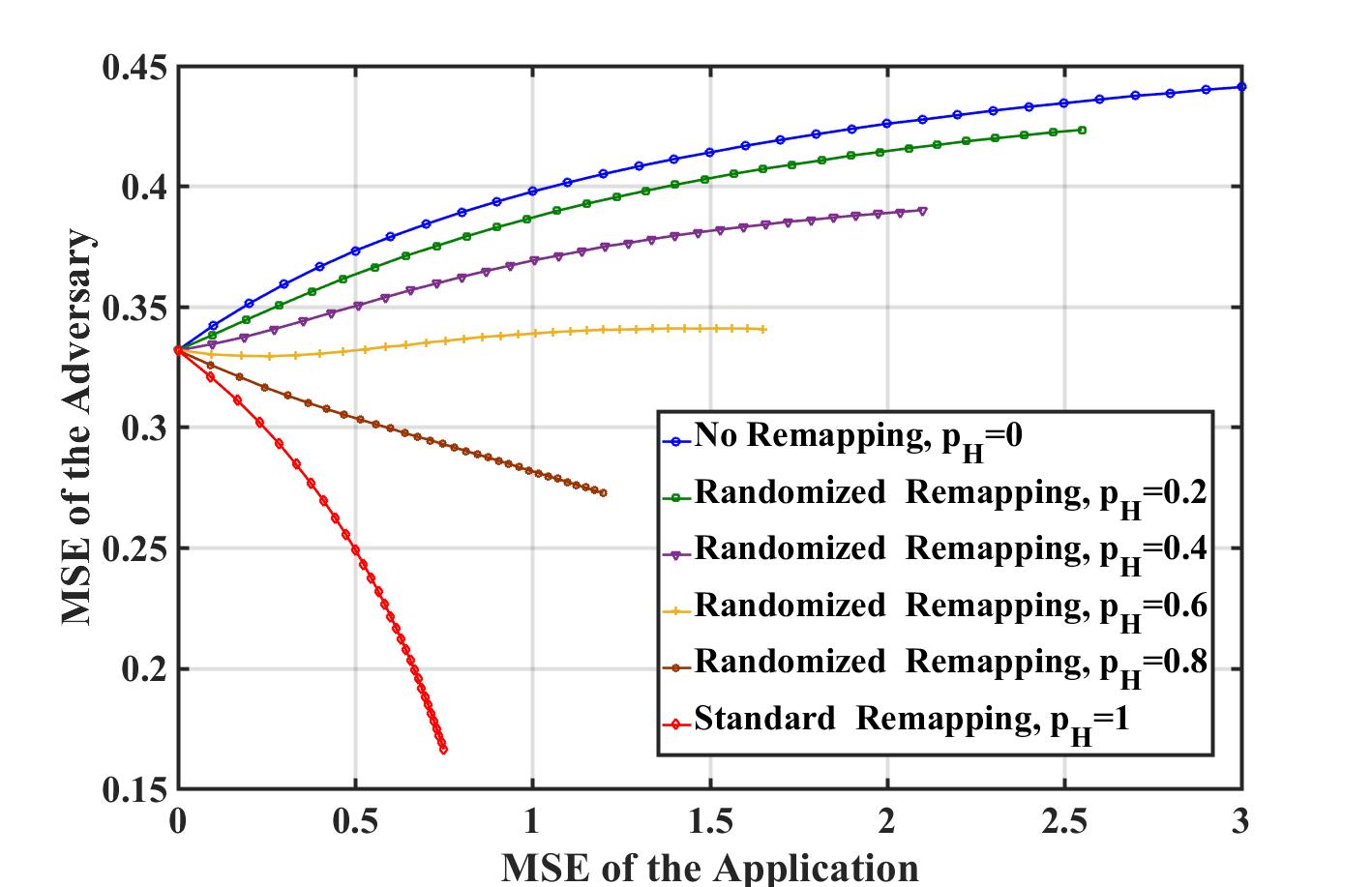}
		\label{fig:randomized}
	}
	\subfloat[The MSE of the adversary about the user' true location $(X)$]{

		\includegraphics[width=0.9\columnwidth]{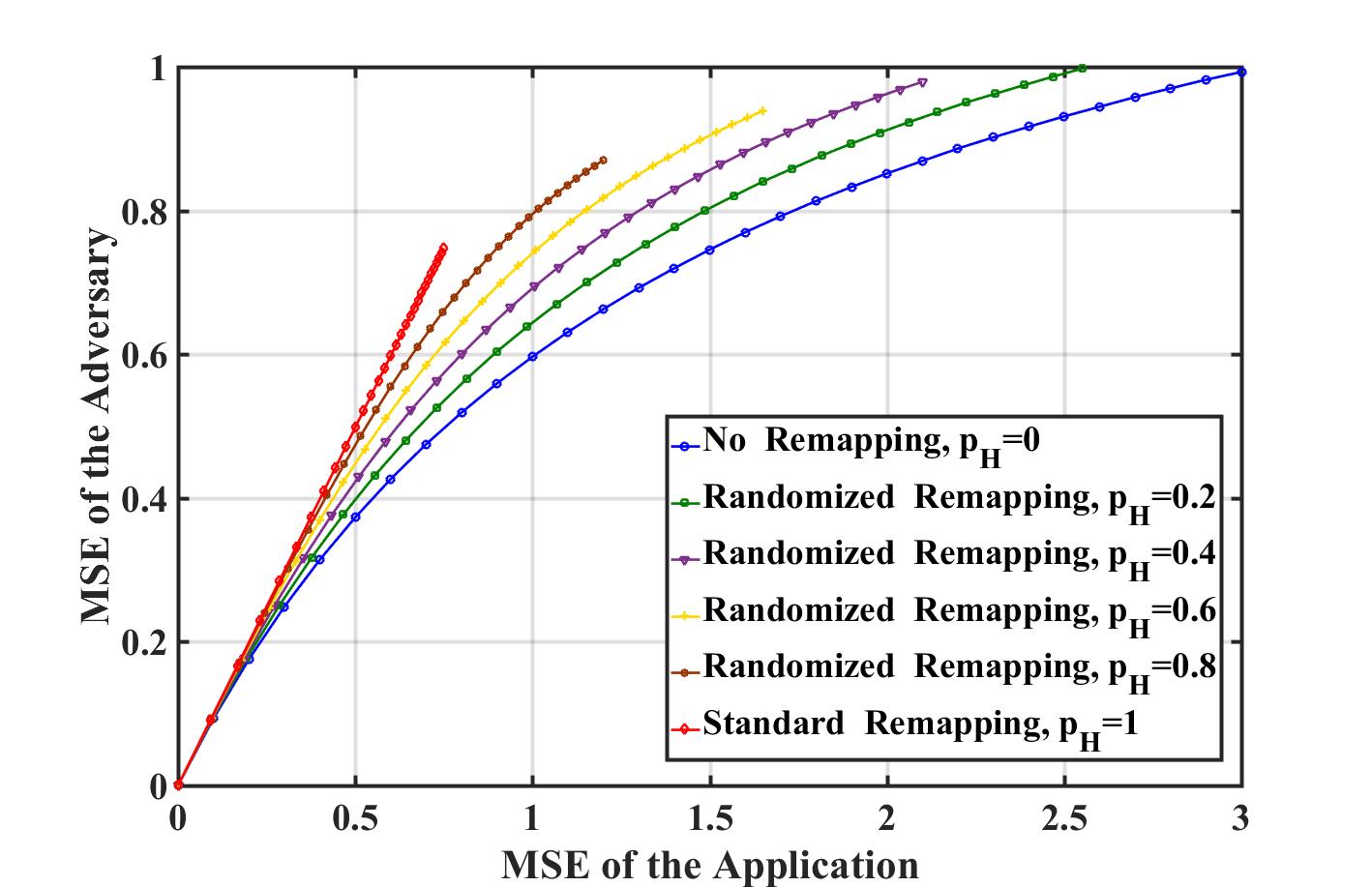}
		\label{fig:randomized2}
	}
	\caption{The MSE of the adversary versus the MSE of the application for three cases. Case 1: remapping technique is not employed ($p_H=0$), Case 2: a randomized remapping technique is employed with $p_H=0.2$, $0.4$, $0.6$, and $0.8$, and Case 3: standard remapping~\cite{remap1} is employed ($p_H=1$). Here, we assume $\sigma^2_\mu=\sigma^2_e=\sigma^2_s=1$ and $\sigma^2_w$ is swept from $0$ to $1$ with steps of $0.1$.}
	\label{fig:random}
\end{figure*}


As derived in Section~\ref{case2}, the remapping technique can leak information about the distribution of the true location $(\mu)$ if the adversary does not have the perfect prior about the user. Here, we introduce a new technique called randomized remapping to improve privacy. This technique provides a trade-off between the leakage of the distribution of the true location $(\mu)$ and the leakage of the true location $({X})$. In the randomized remapping, we have an unfair coin where the probability of a head is equal to $p_H$. For each location, we toss the coin and if a head is observed, the remapped location $(Y_R)$ is released, and if a tail is observed, the noisy version of location $(Y)$ is released. As a result,
\[
{Z}=\begin{cases}
Y_R, & \textrm{with probability } p_{H},\\
Y,& \textrm{with probability of } 1-p_H,
\end{cases}
\]
The user's friend observes $Z$; thus, the MSE of the application is
\begin{align}
\mathcal{U}^{(III)}_{Rand}&=\mathbb{E}\left[\left(Z-X\right)^2\right]=p_H\frac{\sigma_w^2}{\sigma_w^2+\sigma_s^2}+\left(1-p_H\right)\sigma_w^2.
\label{rand_U}
\end{align}
However, the adversary observes both $Z$ and $\check{\mu} =\mu +E$ to estimate the true location $(X)$ and distribution of the true location $(\mu)$. We can calculate $\grave{\mathcal{P}}_{Rand}^{(III)}$ which indicates the MSE of the adversary about the distribution of the true location $(\mu)$ as: 
\begin{align}
\grave{\mathcal{P}}_{Rand}^{(III)}=\mathbb{E}\left[\left(\widetilde{\mu}_{Adv}-\mu\right)^2\right].
\end{align}
Figure~\ref{fig:randomized} shows the MSE of the adversary about the statistical model $(\grave{\mathcal{P}}^{(III)}_{Rand})$ versus the MSE of the intended application/friend $({\mathcal{U}}^{(III)}_{Rand})$.
 We can also calculate ${\mathcal{P}}_{Rand}^{(III)}$ which indicates the MSE of the adversary about the true location $(X)$ as: 
\begin{align}
{\mathcal{P}}_{Rand}^{(III)}=\mathbb{E}\left[\left(\widetilde{X}_{Adv}-X\right)^2\right].
\end{align}
Figure~\ref{fig:randomized2} shows the MSE of the adversary about the true location $(\mathcal{P}^{(III)}_{Rand})$ versus the MSE of the intended application/friend $({\mathcal{U}}^{(III)}_{Rand})$.

From Figures~\ref{fig:randomized} and~\ref{fig:randomized2}, we can conclude that \emph{the standard remapping leaks significant information about the statistical model}, while providing the best privacy level for the user' true location. 
Here, the randomized remapping provides a much better trade-off compared to standard remapping. The value of $p_H$ is a design parameter, so, based on the application requirements and privacy requirements, the appropriate value of $p_H$ should be chosen.



\section{Conclusions}\label{con}
The technique of ``remapping'' has been introduced in the privacy literature to improve the utility of a naive intended recipient while maintaining the same level of privacy against a sophisticated adversary, in particular one with a prior distribution of the user's data. 

We have provided an information-theoretic investigation of this technique. 
We first formulated and analyzed remapping under the standard assumption of an adversary with \textit{perfect} knowledge of the statistical model for the user location. 
Then, we showed that if the adversary has \textit{imperfect} knowledge of the statistical model, the standard remapping technique leaks privacy of both the released location data and the model.
Finally, we proposed a new method called \emph{randomized remapping} which makes it difficult for the adversary to improve their prior knowledge at a given utility, thus providing a better utility-privacy trade-off than the standard remapping. 
Future research will consider an extension of the analysis and countermeasures for generic classes of private datasets.


\bibliographystyle{IEEEtran}
\bibliography{REF}

\begin{thebibliography}{10}
\providecommand{\url}[1]{#1}
\csname url@samestyle\endcsname
\providecommand{\newblock}{\relax}
\providecommand{\bibinfo}[2]{#2}
\providecommand{\BIBentrySTDinterwordspacing}{\spaceskip=0pt\relax}
\providecommand{\BIBentryALTinterwordstretchfactor}{4}
\providecommand{\BIBentryALTinterwordspacing}{\spaceskip=\fontdimen2\font plus
\BIBentryALTinterwordstretchfactor\fontdimen3\font minus
  \fontdimen4\font\relax}
\providecommand{\BIBforeignlanguage}[2]{{%
\expandafter\ifx\csname l@#1\endcsname\relax
\typeout{** WARNING: IEEEtran.bst: No hyphenation pattern has been}%
\typeout{** loaded for the language `#1'. Using the pattern for}%
\typeout{** the default language instead.}%
\else
\language=\csname l@#1\endcsname
\fi
#2}}
\providecommand{\BIBdecl}{\relax}
\BIBdecl

\bibitem{FTC2015}
{Federal Trade Commission Staff}, ``Internet of things: Privacy and security in
  a connected world,'' 2015.

\bibitem{gruteser2003anonymous}
M.~Gruteser and D.~Grunwald, ``Anonymous usage of location-based services
  through spatial and temporal cloaking,'' in \emph{Proceedings of the 1st
  international conference on Mobile systems, applications and services}.\hskip
  1em plus 0.5em minus 0.4em\relax San Francisco, California, USA: ACM, 2003,
  pp. 31--42.

\bibitem{bordenabe2014optimal}
N.~E. Bordenabe, K.~Chatzikokolakis, and C.~Palamidessi, ``Optimal
  geo-indistinguishable mechanisms for location privacy,'' in \emph{Proceedings
  of the 2014 ACM SIGSAC Conference on Computer and Communications
  Security}.\hskip 1em plus 0.5em minus 0.4em\relax Scottsdale, Arizona, USA:
  ACM, 2014, pp. 251--262.

\bibitem{obf_1}
M.~Gruteser and D.~Grunwald, ``Anonymous usage of location-based services
  through spatial and temporal cloaking,'' in \emph{Proceedings of the 1st
  international conference on Mobile systems, applications and services}.\hskip
  1em plus 0.5em minus 0.4em\relax San Francisco, California, USA: ACM, 2003.

\bibitem{obf_3}
C.~Dwork, K.~Kenthapadi, F.~McSherry, I.~Mironov, and M.~Naor, ``Our data,
  ourselves: Privacy via distributed noise generation,'' in \emph{EUROCRYPT},
  2006.

\bibitem{shokri2012protecting}
R.~Shokri, G.~Theodorakopoulos, C.~Troncoso, J.-P. Hubaux, and J.-Y. Le~Boudec,
  ``Protecting location privacy: optimal strategy against localization
  attacks,'' in \emph{Proceedings of the 2012 ACM conference on Computer and
  communications security}.\hskip 1em plus 0.5em minus 0.4em\relax ACM, 2012,
  pp. 617--627.

\bibitem{Nazanin_IT}
N.~Takbiri, A.~Houmansadr, D.~L. Goeckel, and H.~Pishro{-}Nik, ``Matching
  anonymized and obfuscated time series to users' profiles,'' \emph{IEEE
  Transactions on Information Theory}, vol.~65, no.~2, pp. 724--741, 2019.

\bibitem{ISIT18-longversion}
N.~Takbiri, A.~Houmansadr, D.~L. Goeckel, and H.~Pishro{-}Nik, ``Privacy of dependent users against statistical matching,''
  \emph{submitted to IEEE Transactions on Information Theory, Available at
  https://arxiv.org/abs/1710.00197}.

\bibitem{uti_1}
M.~Duckham and L.~Kulik, ``A formal model of obfuscation and negotiation for
  location privacy,'' in \emph{International conference on pervasive
  computing}.\hskip 1em plus 0.5em minus 0.4em\relax Springer, 2005, pp.
  152--170.

\bibitem{remap1}
K.~Chatzikokolakis, E.~ElSalamouny, and C.~Palamidessi, ``Efficient utility
  improvement for location privacy,'' \emph{Proceedings on Privacy Enhancing
  Technologie}, vol. 2017, no.~4, pp. 210--231, 2017.

\bibitem{xiao2005scheme}
L.~Xiao, S.~Boyd, and S.~Lall, ``A scheme for robust distributed sensor fusion
  based on average consensus,'' in \emph{IPSN 2005. Fourth International
  Symposium on Information Processing in Sensor Networks, 2005.}\hskip 1em plus
  0.5em minus 0.4em\relax IEEE, 2005, pp. 63--70.

\bibitem{vempaty2012localization}
A.~Vempaty, O.~Ozdemir, K.~Agrawal, H.~Chen, and P.~K. Varshney, ``Localization
  in wireless sensor networks: Byzantines and mitigation techniques,''
  \emph{IEEE Transactions on Signal Processing}, vol.~61, no.~6, pp.
  1495--1508, 2012.

\bibitem{ribeiro2006bandwidth}
A.~Ribeiro and G.~B. Giannakis, ``Bandwidth-constrained distributed estimation
  for wireless sensor networks-part i: Gaussian case,'' \emph{IEEE transactions
  on signal processing}, vol.~54, no.~3, pp. 1131--1143, 2006.

\bibitem{wagner2004resilient}
D.~Wagner, ``Resilient aggregation in sensor networks,'' in \emph{SASN},
  vol.~4.\hskip 1em plus 0.5em minus 0.4em\relax Citeseer, 2004, pp. 78--87.

\bibitem{gau1}
C.~Cheng, H.~Yang, I.~King, and M.~R. Lyu, ``Fused matrix factorization with
  geographical and social influence in location-based social networks,'' in
  \emph{AAAI}, 2012.

\bibitem{gau2}
Y.~Yu and X.~Chen, ``A survey of point-of-interest recommendation in
  location-based social networks,'' in \emph{AAAI 2015}, 2015.

\bibitem{gau3}
R.~Duan, C.~Jiang, H.~K. Jain, Y.~Ding, and D.~Shu, ``Integrating geographical
  and temporal influences into location recommendation: a method based on
  check-ins,'' \emph{Information Technology and Management}, vol.~20, pp.
  73--90, 2019.

\bibitem{gau4}
Y.~Lyu, C.-Y. Chow, R.~Wang, and V.~C.~S. Lee, ``imcrec: A multi-criteria
  framework for personalized point-of-interest recommendations,'' \emph{Inf.
  Sci.}, vol. 483, pp. 294--312, 2019.

\bibitem{gau5}
S.~Zhao, I.~King, and M.~R. Lyu, ``Capturing geographical influence in poi
  recommendations,'' in \emph{ICONIP}, 2013.

\bibitem{Palia2017OptimizingNL}
A.~Palia and R.~Tandon, ``Optimizing noise level for perturbing geo-location
  data,'' \emph{Future of Information and Communication Conference}, pp.
  63--73, 2018.

\bibitem{rem2}
S.~Oya, C.~Troncoso, and F.~P{\'e}rez-Gonz{\'a}lez, ``Is
  geo-indistinguishability what you are looking for?'' in \emph{Proceedings of
  the 2017 on Workshop on Privacy in the Electronic Society}.\hskip 1em plus
  0.5em minus 0.4em\relax Dallas, Texas, USA: ACM, 2017, pp. 137--140.

\bibitem{rem3}
R.~Mendes and J.~P. Vilela, ``On the effect of update frequency on
  geo-indistinguishability of mobility traces,'' in \emph{Proceedings of the
  11th ACM Conference on Security and Privacy in Wireless and Mobile
  Networks}.\hskip 1em plus 0.5em minus 0.4em\relax Stockholm, Sweden: ACM,
  2018, pp. 271--276.

\bibitem{rem4}
\BIBentryALTinterwordspacing
Y.~Kawamoto and T.~Murakami, ``Differentially private obfuscation mechanisms
  for hiding probability distributions,'' \emph{CoRR}, vol. abs/1812.00939,
  2018. [Online]. Available: \url{http://arxiv.org/abs/1812.00939}
\BIBentrySTDinterwordspacing

\bibitem{rem5}
\BIBentryALTinterwordspacing
S.~Oya, C.~Troncoso, and F.~P{\'{e}}rez{-}Gonz{\'{a}}lez, ``A tabula rasa
  approach to sporadic location privacy,'' \emph{CoRR}, vol. abs/1809.04415,
  2018. [Online]. Available: \url{http://arxiv.org/abs/1809.04415}
\BIBentrySTDinterwordspacing

\bibitem{ISIT2019_longversion}
N.~Takbiri, A.~Houmansadr, D.~L. Goeckel, and H.~Pishro{-}Nik, ``Remapping,''
  \url{http://www.ecs.umass.edu/ece/pishro/Papers/remapping.pdf}, January,.

\bibitem{noise1}
A.~Solanas and A.~Mart{\'i}nez-Ballest{\'e}, ``Privacy protection in
  location-based services through a public-key privacy homomorphism,'' in
  \emph{EuroPKI}.\hskip 1em plus 0.5em minus 0.4em\relax Springer, 2007.

\bibitem{noise2}
A.~Solanas, J.~Domingo-Ferrer, and A.~Mart{\'i}nez-Ballest{\'e}, ``Location
  privacy in location-based services: Beyond ttp-based schemes,'' in
  \emph{PiLBA}, 2008.

\bibitem{noise3}
J.~Krumm, ``Inference attacks on location tracks,'' in \emph{Pervasive}, 2007.

\bibitem{randomizedresponse}
S.~L. Warner, ``Randomized response: A survey technique for eliminating evasive
  answer bias,'' \emph{Journal of the American Statistical Association},
  vol.~60, no. 309, pp. 63--69, 1965.

\end{thebibliography}
%
%
%
%

\end{document}